\documentclass[prl,aps,twocolumn,amsmath,amssymb,floatfix]{revtex4}

\usepackage{amsmath}
\usepackage{graphics}
\usepackage{graphicx}
\usepackage{rotating}

\begin{document}

\title{Graphene to Graphane: The Role of H Frustration in Lattice Contraction}

\author{S. B. Legoas$^{1}$}
\author{P. A. S. Autreto$^{2}$}
\author{M. Z. S. Flores$^{2}$}
\author{D. S. Galv\~{a}o$^{2}$}
 \email{galvao@ifi.unicamp.br}
 
\affiliation{$^{1}$Centro de Ci\^{e}ncias e Tecnologia, Universidade Federal de Roraima, 69304-000, Boa Vista, Roraima, Brazil}

\affiliation{$^{2}$Instituto de F\'{i}sica ``Gleb Wataghin'', Universidade Estadual de Campinas, Unicamp, C.P. 6165, 13083-970, Campinas, S\~{a}o Paulo, Brazil}

\date{February, 16th, 2009}

\begin{abstract}
Graphane is a two-dimensional system consisting of a single planar layer of fully saturated (sp$^3$ hybridization) carbon atoms with H atoms attached to them in an alternating pattern (up and down with relation to the plane defined by the carbon atoms). Stable graphane structures were theoretically predicted to exist some years ago and just experimentally realized through hydrogenation of graphene membranes. In this work we have investigated  using \textit{ab initio} and reactive molecular dynamics the role of H frustration (breaking the H atoms up and down  alternating pattern) in graphane-like structures. Our results show that H frustration significantly contributes to lattice contraction. The dynamical aspects of converting graphene to graphane is also addressed.
\end{abstract}



\maketitle

Carbon-based materials are among the most studied ones, however the discovery of new structures seems endless, colossal carbon tubes \cite{colossal} and graphene \cite{geim} being recent examples.

Graphenes are flat monolayers of carbon atoms in a sp$^2$ hybridization. They have very interesting electronic and mechanical properties, which make them one of most important subject in materials science today \cite{geim,geim1,castroneto}.

It has been theoretically predicted that a related structure, called graphane \cite{sofo}, could exist. Graphane consists of a single planar layer structure with fully saturated (sp$^3$ hybridization) carbon atoms with H attached to them in an alternating pattern (up and down with relation to the plane defined by the carbon atoms). The two most stable conformations are the so-called chair-like (H atoms alternating on both sides of the plane) and boat-like (H atoms alternating in pairs) \cite{sofo} (Figure 1). A third member of these two-dimensional planar carbon structures, called graphynes \cite{ray1,ray2,ray3}, have also been predicted to exist but up to now only molecular fragments have been synthesized \cite{ray2}.

\begin{figure}[b]
\begin{center}
\includegraphics[scale=0.4]{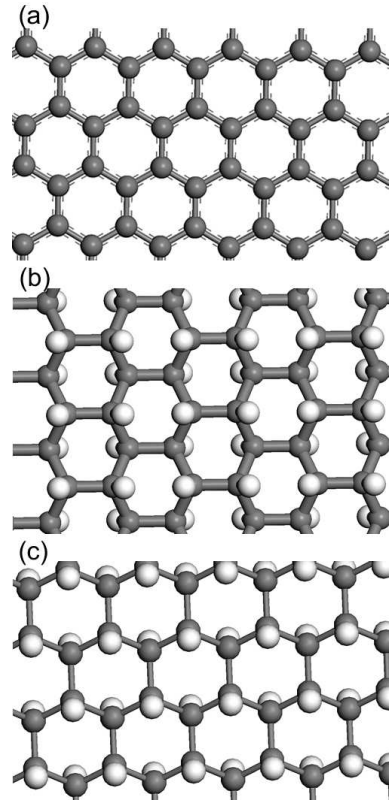}
\caption{Structural models. (a) graphene; (b) graphane-boatlike; (c) graphane-chairlike. See text for discussions.}
\label{figure1}
\end{center}
\end{figure}

Indirect experimental evidences of graphane-like structures have been reported \cite{humberto,graphane2}. More recently, in a series of very elegant experiments, Elias \textit{et al.} \cite{geim2} demonstrated the existence of graphane formation from graphene through hydrogenation with cold plasma. They also demonstrated that the process is reversible. These fundamental discoveries open new and important perspectives to the use of graphene-based devices since the electronic gap values in graphanes could be controlled by the degree of hydrogenation \cite{geim2,comment}.

The Elias \textit{et al.} experiments consisted in exposing graphene membranes to H$^{+}$ from cold plasma. The H incorporation into the membranes results in altering the C sp$^2$ hybridizations to sp$^3$ ones. The experiments were also done with the membranes over SiO$_2$ substrates (only one membrane side exposed to H$^{+}$ attacks) and produced a material with different properties.

One very interesting result from their TEM (transmission electron microscopy) data is the distribution of the lattice spacing value of the hydrogenated membranes. For an ideal graphane structure theoretical calculations indicate a lattice spacing value larger than the graphene one \cite{sofo}. Although the TEM data (Fig. 3 from \cite{geim2}) show events with values larger than the graphene one, the majority of counts show smaller ones.

Elias \textit{et al.} discuss in details a series of experimental and material conditions that could explain the contracted lattice values, but from the experimental point of view it is very difficult to identify and/or isolate specific contributions to this effect. Of particular interest it is the role of the ordered and/or disordered arrays of H incorporation into the membranes.

In this work we have investigated using \textit{ab initio} quantum and classical binding energy bond order (BEBO) molecular dynamics methods how the graphene hydrogenation is reflected into lattice deformations.

We have carried out \textit{ab initio} total energy calculations in the framework of the density functional theory (DFT), as implemented in the DMol$^3$ code \cite{dmol3}. DMol$^3$ is considered state of the art DFT methodology. Exchange and correlation terms were treated within the generalized gradient (GGA) functional by Perdew, Burke, and Ernzerhof \cite{pbe}. Core electrons were treated in a non-relativistic all electron implementation of the potential. A double numerical quality  basis set  with polarization function (DNP) were considered, with a real space cutoff of 3.7 \AA\ . The tolerances of energy, gradient, and displacement convergence were 0.00027 eV, 0.054 eV/\AA\, and 0.005 \AA\, respectively.

We have considered finite and infinite (cyclic boundary conditions(CBC)) structures. As we need to investigate many structures of different sizes and at different temperatures (molecular dynamics (MD) simulations) the extensive use of \textit{ab initio} methods is computationally cost prohibitive. Thus, for the larger structures and for MD simulations, we opted to use a BEBO method like ReaxFF \cite{duin1,duin2,duin3}. ReaxFF is a reactive force field developed by Adri van Duin, William Goddard III and co-workers for use in MD simulations. This empirical method allows the simulation of many types of chemical reactions, including bond dissociation (making/breaking bonds).

ReaxFF is similar to standard non-reactive force fields, like MM3 \cite{mm3}, where the system energy is divided into partial energy contributions associated with, amongst others, valence angle bending, bond stretching, and non-bonded van der Waals and Coulomb interactions \cite{duin1,duin2}. However, one main difference it is that ReaxFF can handle bond formation and dissociation using bond order approach. ReaxFF was parameterized against DFT calculations. The average deviation between the ReaxFF predicted heat of formation values and the experimental ones are of 2.8 and 2.9 Kcal/mol for non-conjugated and conjugated systems, respectively \cite{duin2}.

We have carried out geometry optimizations using gradient conjugated techniques (stopping condition, gradient values less than 10$^-$$^3$). For MD simulations we used scaling methods, like Beredsen thermostat \cite{berendsen}, in order to control temperature.

We started carrying out DMol$^3$ calculations for the infinite (CBC) structures shown in Fig. 1. The cell unit parameters were optimized using the Murnaghan \cite{murnaghan} procedure. Sofo, Chaudhari, and Baker \cite{sofo} in their pioneering graphane work considered compacted (interacting layers) structures. Here, in order to mimic the experimental conditions \cite{geim2}, we have considered isolated (non-interacting) layers. The results are displayed in table I. The most stable conformation (chair) has just one type of C-C bonds, while the boat conformation has two types (carbon bonded to H on opposite side of plane of symmetry and those connected to H on the same side of it), with their associated H bonds.

\begin{table}[h]
\centering
\caption{DMol$^3$ results for the crystalline structures shown in Fig. 1. See text for discussions.}
\begin{tabular}{llll}
\hline
\bf{System} & \multicolumn{1}{c}{\bf{Energy(Ha)}} & \multicolumn{1}{c}{\bf{
Lattice(\AA)}} & \multicolumn{1}{c}{\bf{C-C(\AA)}} \\ 
\hline
Graphene  & \multicolumn{1}{c}{-304.68} & \multicolumn{1}{c}{2.465} & \multicolumn{1}{c}{1.423} \\ 
G-chair & \multicolumn{1}{c}{-309.41} & \multicolumn{1}{c}{2.540} & \multicolumn{1}{c}{1.537} \\ 
G-boat  & \multicolumn{1}{c}{-309.38} & \multicolumn{1}{c}{2.592} & \multicolumn{1}{c}{1.581} \\ 
        &                             & \multicolumn{1}{c}{2.509} & \multicolumn{1}{c}{1.537} \\
\hline
\end{tabular}
\label{1st_table}
\end{table}

\begin{figure}[b]
\begin{center}
\includegraphics[width=6.5cm]{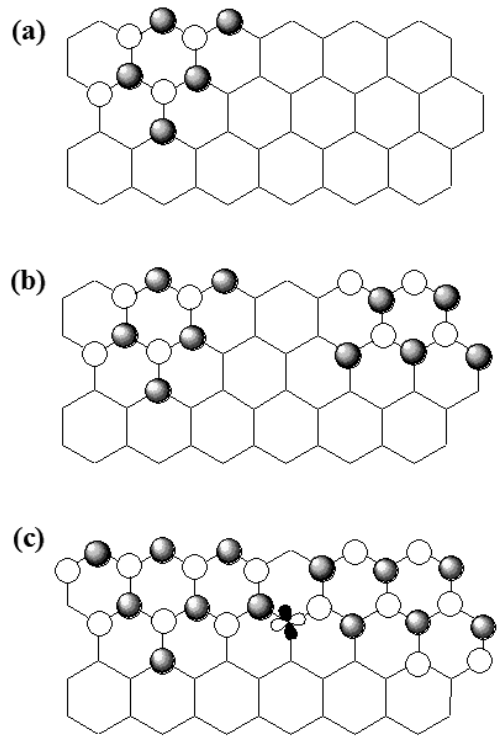}
\caption{Scheme of the H frustrated domains. See text for discussions.}
\label{Pt_tip}
\end{center}
\end{figure}

As we can see from Table I, the lattice parameter values for the graphane structures are larger than the graphene one. We have also considered the case of the minimum unit cell with H atoms parallelly aligned (just one side of the membrane). Our results show that the system is unstable with the tendency of H$_2$ recombination and/or C-C breaking bonds. These results are in good agreement with previous work \cite{sofo} and with the available experimental data \cite{geim,geim2}.
 
Considering the stochastic nature of the experiments, there is a significant probability of the existence of H frustration (Fig. 2). H frustration is a configuration where the sequence of alternating up and down H atoms is broken (frustrated). This is similar to spin frustration in magnetic materials \cite{spin}. 

In Fig. 2a we show a domain of up and down H atoms. From energetic (stability) point of view, after the first (up or down) H is incorporated into the C layer, the next favorable site is its first inverse neighbor (down or up), and so on. If the system is large enough uncorrelated domains might be formed (Fig. 2b). As the H coverage is continued it could occur that it is no longer possible the alternating sequence of up and down H atoms (Fig. 2c).

We have investigated finite fragments with and without H frustrations. We analyzed the associated geometrical changes (Fig. 3) in order to determine whether the lattice expands or contracts with relation to the ideal graphane structure (table I). We have carried out DMol$^3$ (table II) and ReaxFF (for larger fragments) calculations (table III, Fig. 4).

\begin{figure}[h]
\begin{center}
\includegraphics[scale=0.4]{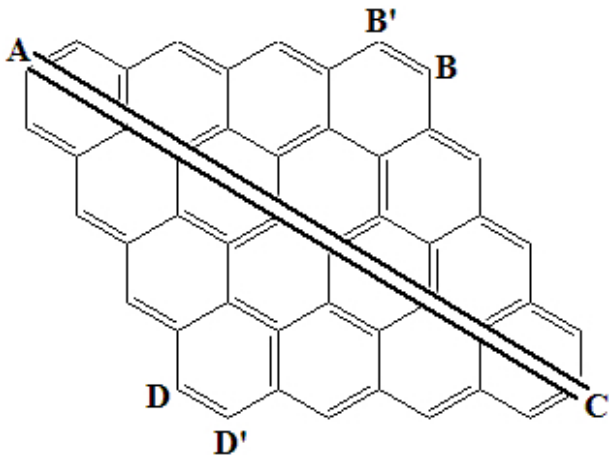}
\caption{Schematic draw of a graphene fragment before hydrogenation. The letters are the reference points for the distances displayed in Tables II and III. }
\label{twin}
\end{center}
\end{figure}

\begin{figure}[b]
\begin{center}
\includegraphics[scale=0.42]{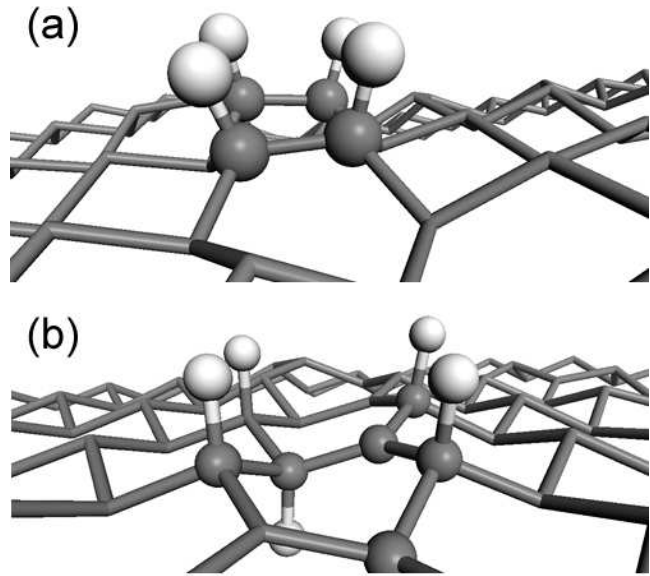}
\caption{Examples of the optimized geometries for some of the frustrated structures listed in tables II and III. (a) frust-1, H frustration with parallel first-neighbor H atoms; (b) frust-0, H frustration with 
'missing' first-neighbor H atoms. Atoms in the defect region are shown in ball and stick rendering. For view clarity the H atoms outside this region were made transparent.} 
\label{modelo}
\end{center}
\end{figure}

DMol$^3$ and ReaxFF show similar and consistent results. The H frustration produce out of plane distortions which induce in-plane shrinkage. The net result is a decrease of the lattice parameter in relation to the ideal graphane value. This effect is amplified when first-neighbor H atoms are parallelly aligned (Fig. 4). 

\begin{table}[h]
\centering
\caption{Distances between reference points depicted in Fig. 3. Frust-1 and Frust-0 refer to parallel (see Fig. 4a) and missing (see Fig. 4b) hydrogen atoms in frustrated domains, respectively.}
\begin{tabular}{lllllll}
\hline
\bf{System} & \multicolumn{1}{c}{\bf{d$_{A-B}$}} & \multicolumn{1}{c}{\bf{d$_{B'-C}$}} & \multicolumn{1}{c}{\bf{d$_{C-D}$}} & \multicolumn{1}{c}{\bf{d$_{D'-A}$}} & \multicolumn{1}{c}{\bf{d$_{A-C}$}} & \multicolumn{1}{c}{\bf{d$_{B-D}$}} \\ 
\hline
Graphene & \multicolumn{1}{c}{9.804} & \multicolumn{1}{c}{9.799} & \multicolumn{1}{c}{9.804} & \multicolumn{1}{c}{9.799} & \multicolumn{1}{c}{15.569} & \multicolumn{1}{c}{9.974} \\ 
G-chair & \multicolumn{1}{c}{9.861} & \multicolumn{1}{c}{9.841} & \multicolumn{1}{c}{9.882} & \multicolumn{1}{c}{9.847} & \multicolumn{1}{c}{15.636} & \multicolumn{1}{c}{10.050} \\ 
G-boat & \multicolumn{1}{c}{9.852} & \multicolumn{1}{c}{9.818} & \multicolumn{1}{c}{9.852} & \multicolumn{1}{c}{9.818} & \multicolumn{1}{c}{15.622} & \multicolumn{1}{c}{9.977} \\ 
frust-0 & \multicolumn{1}{c}{9.788} & \multicolumn{1}{c}{9.857} & \multicolumn{1}{c}{9.876} & \multicolumn{1}{c}{9.823} & \multicolumn{1}{c}{15.588} & \multicolumn{1}{c}{10.003} \\ 
frust-1 & \multicolumn{1}{c}{9.740} & \multicolumn{1}{c}{9.802} & \multicolumn{1}{c}{9.866} & \multicolumn{1}{c}{9.786} & \multicolumn{1}{c}{15.465} & \multicolumn{1}{c}{9.990} \\ 
\hline
\end{tabular}
\label{2nd_table}
\end{table}

\begin{table}[h]
\centering
\caption{Distances between the reference points depicted in Fig. 3 calculated with ReaxFF. The number in parenthesis indicate the number of frustrated domains in the structure. See text for discussions.}
\begin{tabular}{lllllll}
\hline
\bf{System} & \multicolumn{1}{c}{\bf{d$_{A-B}$}} & \multicolumn{1}{c}{\bf{d$_{B'-C}$}} & \multicolumn{1}{c}{\bf{d$_{C-D}$}} & \multicolumn{1}{c}{\bf{d$_{D'-A}$}} & \multicolumn{1}{c}{\bf{d$_{A-C}$}} & \multicolumn{1}{c}{\bf{d$_{B-D}$}} \\ 
\hline
Graph. & \multicolumn{1}{c}{22.26} & \multicolumn{1}{c}{22.26} & \multicolumn{1}{c}{22.26} & \multicolumn{1}{c}{22.26} & \multicolumn{1}{c}{37.08} & \multicolumn{1}{c}{22.47} \\ 
G-chair & \multicolumn{1}{c}{22.95} & \multicolumn{1}{c}{22.94} & \multicolumn{1}{c}{22.95} & \multicolumn{1}{c}{22.94} & \multicolumn{1}{c}{38.40} & \multicolumn{1}{c}{23.01} \\ 
Frust-1 (13) & \multicolumn{1}{c}{22.21} & \multicolumn{1}{c}{22.92} & \multicolumn{1}{c}{22.80} & \multicolumn{1}{c}{22.90} & \multicolumn{1}{c}{37.95} & \multicolumn{1}{c}{22.47} \\ 
Frust-0 (13) & \multicolumn{1}{c}{22.38} & \multicolumn{1}{c}{22.95} & \multicolumn{1}{c}{22.77} & \multicolumn{1}{c}{22.82} & \multicolumn{1}{c}{38.12} & \multicolumn{1}{c}{22.37} \\ 
\hline
\end{tabular}
\label{Pedro_table}
\end{table}

\begin{figure}[t]
\begin{center}
\includegraphics[scale=0.42]{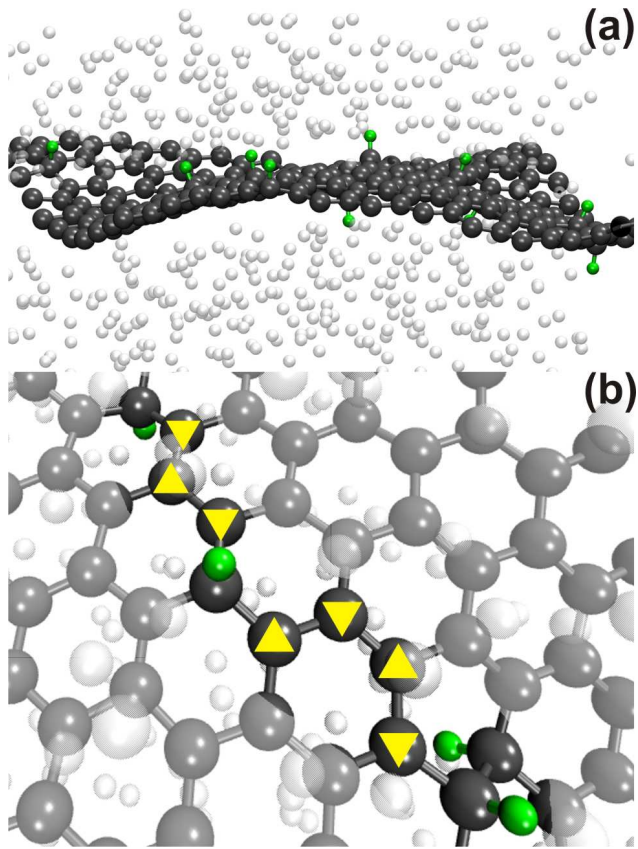}
\caption{(a) Representative snapshot of the early hydrogenation stages from ReaxFF molecular simulations at 300 K. Free atomic H atoms are indicated in white and C-bonded ones in green. (b) zoomed region indicating H frustrated domains formed. The triangle path shows that a sequence of up and down H atoms is no longer possible.}
\label{modelo}
\end{center}
\end{figure}

In order to further test these ideas we have also carried out ReaxFF molecular dynamics simulations of graphene hydrogenation on very large membranes, at different temperatures and at different H atom densities.

In Fig. 5 we show representative snapshots from the early stages of a simulation at 300 K. The results show that significant percentage of uncorrelated H frustrated domains are formed in the early stages of the hydrogenation process leading to lattice decreased values and extensive membrane corrugations. These results also suggest that large domains of perfect graphane-like structures are unlikely to be formed, H frustrated domains are always present. The number of these domains seems to be sensitive to small variations of temperatures and H gas densities. This can perhaps explain the significant broad lattice parameter distribution values experimentally observed \cite{geim2}. 

We run annealing cycle simulations to verify whether thermal healing occurs. Our preliminary results show that, once formed, H frustrated domain are resilient to thermal annealing. Further studies are necessary to clarify this issue.

A detailed MD analysis including the effect of many layers, substrates and association of inert gas with H atoms will be published elsewhere \cite{graphane3}.

Acknowledgments - This work was supported in part by the Brazilian Agencies CNPq, CAPES and FAPESP. The authors wish to thank Prof. A. van Duin for his very helpful assistance with the ReaxFF code and for his kind hospitality in Pennsylvania.

\end{document}